# An oxygen-rich, tetrahedral surface phase on high-temperature rutile VO$_2$(110)$_T$ single crystals

Margareta Wagner[1*], Jakub Planer[1,2], Bettina S. J. Heller[3], Jens Langer[4], Andreas Limbeck[5], Lynn A. Boatner[6], Hans-Peter Steinrück[3], Josef Redinger[1,2], Florian Maier[3], Florian Mittendorfer[1,2], Michael Schmid[1] and Ulrike Diebold[1]

[1] *Institute of Applied Physics, TU Wien, Wiedner Hauptstraße 8–10/134, 1040 Vienna, Austria*
[2] *Center for Computational Materials Science, TU Wien, Wiedner Hauptstraße 8–10/134, A-1040 Vienna, Austria*
[3] *Chair of Physical Chemistry II, Friedrich-Alexander-Universität Erlangen-Nürnberg (FAU), Egerlandstrasse 3, 91058, Erlangen, Germany*
[4] *Chair of Inorganic and Organometallic Chemistry, Friedrich-Alexander-Universität Erlangen-Nürnberg (FAU), Egerlandstrasse 1, 91058 Erlangen, Germany*
[5] *Institute of Chemical Technologies and Analytics, TU Wien, Getreidemarkt 9/164, 1060 Vienna, Austria*
[6] *Materials Science and Technology Division, Oak Ridge National Laboratory, Oak Ridge, Tennessee 37831, USA*

**Abstract**

Vanadium dioxide undergoes a metal-insulator transition (MIT) from an insulating (monoclinic) to a metallic (tetragonal) phase close to room temperature, which makes it a promising functional material for many applications, *e.g.* as chemical sensors. Not much is known about its surface and interface properties, although these are critical in many of its applications. This work presents an atomic-scale investigation of the tetragonal rutile VO$_2$(110)$_T$ single-crystal surface and reports results obtained with scanning tunneling microscopy (STM), low-energy electron diffraction (LEED) and X-ray photoelectron spectroscopy (XPS), supported by density-functional theory-based (DFT) calculations. The surface reconstructs into an oxygen-rich (2×2) superstructure that coexists with small patches of the underlying, unreconstructed (110)-(1×1) surface. The best structural model for the (2×2) surface termination, conceptually derived from a vanadium pentoxide (001) monolayer, consists of rings of corner-shared tetrahedra. Over a wide range of oxygen chemical potentials this reconstruction is more stable than the unreconstructed (110) surface as well as models proposed in the literature.

---

[*] Corresponding author: wagner@iap.tuwien.ac.at

# 1. Introduction

Vanadium(IV) dioxide, $VO_2$, undergoes a first-order metal-to-insulator transition (MIT) at a critical temperature $T_C$ of ~340 K (67°C), where the lattice changes from the monoclinic structure M1 of the semiconducting/insulating phase ($T<T_C$, distorted rutile, space group $P2_1/C$) into a tetragonal (rutile) structure of the metallic phase ($T>T_C$, space group $P4_2/mnm$) [1]. The MIT shows a hysteresis of several K in heating-and-cooling cycles. During the phase transition, an intermediate strain- or doping-induced monoclinic phase M2 has been observed, together with a metastable triclinic phase T occurring between M1 and M2 [2,3]. The physical mechanism behind the complex phase transition of this strongly correlated oxide is still controversially discussed; an overview is given in Ref. [4].

$VO_2$ is technologically interesting as its MIT occurs near room temperature, and the change in resistivity by several orders of magnitude is accompanied by changes in optical (NIR transmittance), thermal and magnetic properties. The MIT can be tailored to ultrafast switching in the range of ~500 femtoseconds [5,6], and, recently, ~26 fs were reported by Jager et al. [7]. Moreover [4], lattice strain [8], also induced by cation or hydrogen doping [9-11], electric current, electric field gating, and irradiation with light [12] modify the MIT and shift its critical temperature even closer to room temperature. The properties and tuneability of the MIT in $VO_2$ are employed in applications, *e.g.*, in memristive devices [13], optical modulators [14], gas sensors [15], field-effect transistors [16] or smart window coatings [17,18].

Since surface properties play an important role in many of these applications, several recent studies have characterized $VO_2$ surfaces, predominantly using supported thin and ultrathin films. It was revealed [19,20] that tensile (compressive) strain along the rutile *c*-axis imposed by a lattice mismatch with the substrate is correlated with an increase (decrease) in transition temperature. Recent evidence points towards a selvedge that is not just simply a bulk termination: Thin $VO_2(110)$ films grown on $RuO_2(110)$ and $TiO_2(110)$ substrates show an oxygen-rich (2×2) surface termination [21,22]. This is in agreement with earlier studies, which indicated an increased concentration of oxygen atoms at the surface not only under ambient conditions [23], but also under reducing conditions comparable to ultrahigh vacuum (UHV) [24]. Moreover, DFT calculations [22] revealed that rutile surfaces are lower in energy compared to the monoclinic counterparts, and also that oxygen-rich reconstructions reduce the occupation of surface 3d states which is an important driving parameter for the MIT [25]. This is in agreement with recent experimental studies [22,26], which did not reveal any evidence of the bulk structural transition to the monoclinic phase at the surface.

These recent results call for a detailed knowledge of the atomic-scale structure of $VO_2$ surfaces. Such information can best be gained by applying complementary surface science techniques to single crystals and combining these experiments with the DFT calculations. This work focuses on the $(110)_T$ (tetragonal) surface of $VO_2$ above $T_C$, where the high electrical conductivity enables such experiments. It starts with a thorough bulk characterization of the structure and purity of the $VO_2$ single crystals, grown from the melt. The experiments were quite challenging; the MIT is connected with a massive structural change. The rigid sample mount required for the surface science measurements can easily lead to a fracture of the samples during the phase transition; to avoid this, the crystals were kept at elevated temperature throughout most of the experiments.

LEED confirms the (2×2) periodicity observed in earlier works [21,22]. X-ray photoemission spectroscopy (XPS) supports that the surface is a vanadium oxide phase enriched with oxygen compared to the bulk and excludes a possible segregation of impurities. Atomically resolved scanning tunneling microscopy (STM) studies show that the (2×2) structure consists of an adlayer, i.e., a modified layer at the very top of the crystal. A detailed analysis of these results constrains the possible structural models.

From the DFT perspective, vanadium dioxide is a challenging material due to strong electron-electron correlations. Benchmark calculations show that standard DFT functionals cannot correctly describe bulk properties like the electron density, relative phase stability, band gap and magnetic ordering for both the rutile and monoclinic phases at the same time [27,28]. Nevertheless, DFT was successfully employed to characterize the surface structure of several $VO_x$-derived surfaces. For example, Schoiswohl *et al.* [29] characterized the formation of ultrathin $VO_x$ structures on a metallic substrate with a combination of atomically resolved STM and *ab initio* calculations, which provided additional information such as the exact stoichiometry or the atomistic structure. Klein *et al.* [30] investigated $V_2O_3$ and $V_5O_{14}$ structures on Pd(111) and confirmed the structures obtained with DFT experimentally by LEED-I(V).

DFT calculations were also employed in the structural and energetic characterization of $V_2O_5$ and $V_6O_{13}$(001) surfaces [31].

In this work, DFT is used to interpret the experimental observations and a novel, tetrahedrally-coordinated surface phase is the best fit to the experimental results.

## 2. Materials and Methods

***Samples*:** The $VO_2$ single crystals were grown from melt using $V_2O_5$ powder kept in an Ar flow at 1000 °C for 120 h in a quartz ($SiO_2$) crucible. The solid black crystals are shaped as needles or thin plates of approximately 2–10 mm length, 1–4 mm in width and <1 mm in thickness, exposing a flat and reflective top side; see Figure 1(a). Larger crystals often consist of several smaller needles grown together. The crystals are brittle and break easily during mounting. Figure 1(b) shows a crystal tightly mounted on a Ta sample plate with a single Ta spring (1 mm wide) for STM measurements. A thin Au foil is placed between the crystal and sample plate to improve thermal contact to the rough back side of the crystal. The crystal broke into several pieces after a week of measurements and sample transfers inside the UHV chamber. Panel (b) shows one of the larger crystals prepared for XPS measurements. It is placed inside a recess on a molybdenum sample holder; the crystal is gently held by a 0.2 mm Chromel wire spot-welded to the edges of the recess. A Chromel-alumel (type K) thermocouple is connected to the crystal.

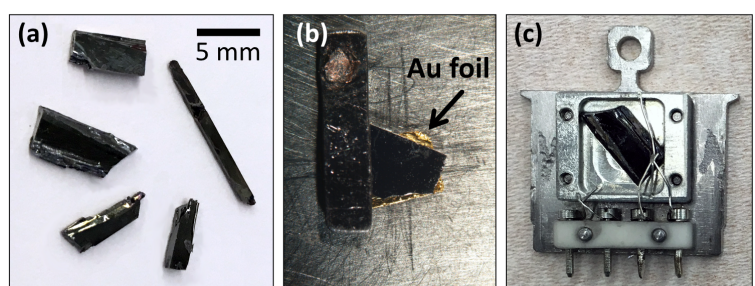

Figure 1: The $VO_2$ single crystals. (a) A selection of large, as-grown crystals, exposing the flat and shiny side. Small crystals are usually needle-shaped, similar to the rightmost crystal in the picture. (b) A crystal mounted for STM measurements, held tightly by a tantalum spring and gold foil. (c) A crystal mounted in a heatable molybdenum sample holder for XPS measurements, with a thermocouple on top and held in place by an additional Chromel wire.

***Bulk impurities*:** The purity of the crystals was investigated by trace analysis with laser ablation inductively coupled plasma mass spectrometry (LA-ICP-MS) using NIST612 for quantification and $^{51}V$ as internal standard. All crystals show a constant $^{51}V$ signal across the whole surface (1–2 mm) and traces of Ba (<2.3 ppm), Pb (< 7 ppm), U (0.25–41 ppm), Mg (<0.42 ppm), Al (0.03–81 ppm), Ta (0.03–302 ppm), and Au (0.0–58 ppm) (see Supplemental Material). The impurities are different for each crystal and are also inhomogeneously distributed within the individual crystals. The Au and Ta signals were particularly high on a crystal that had been previously sputtered in UHV, mounted with a Ta spring on a Ta sample plate with an Au foil underneath the sample (Fig. 1b).

***Structural Characterization***: X-Ray diffraction (XRD) was done with a SuperNova, Dual, Cu at zero, AtlasS2 diffractometer (Rigaku Oxford Diffraction). The purpose of these measurements was not only to confirm the bulk structures above and below the transition temperature, but also to determine the surface orientation of the shiny crystal side. Since this machine requires small samples, it was necessary to cut the original $VO_2$ single crystals into appropriate pieces. Several crystals were investigated, both as-grown and after the XPS measurements (discussed below), which included crossing the phase transition a few times and heating in UHV to 650 °C. As expected, the treatment in UHV does not influence the bulk structure, although it introduced twin formation. In order to keep track of the prominent shiny side of the crystals, which was used in the other experiments, this side was colored using a white varnish before cutting. This made it possible to identify this facet even in case of irregularly shaped fragments. A selected fragment is shown in Fig. 2(a,b) (crystal dimensions in mm: min = 0.08, mid = 0.18, max = 0.35) mounted on a Hampton Research CryoLoop covered with a thin film of inert perfluoropolyalkylether (viscosity 1800 cSt; ABCR GmbH) and investigated at 25 °C and 110 °C, respectively, using Mo Kα radiation ($\lambda = 0.71073$ Å). The temperature during the experiments was kept constant by a Cryostream 800 cooler (Oxford Cryosystems). The data were processed with the CrysAlisPro (v38.46) software package [32], and unit cell determination confirmed the presence of the known monoclinic modification at 25 °C, while the tetragonal form was present at 110 °C. The faces of the crystal were indexed using the Crystal Shape Tools implemented in CrysAlisPro.

In addition to the study of the crystal shown in Fig, 2(a,b), Laue back-reflection analysis was applied to some crystals prior to the STM measurements employing a Siemens Kristalloflex instrument in air and at room temperature with wavelengths from 0.25 to 2.50 Å (30 kV, 18–20 mA, 10 min measurement time). The crystals were first aligned perpendicular to the incident beam/X-Ray gun and afterwards tilted by ~4° to shift the central diffraction spots onto the detector. The diffraction pattern was simulated using the software OrientExpress.

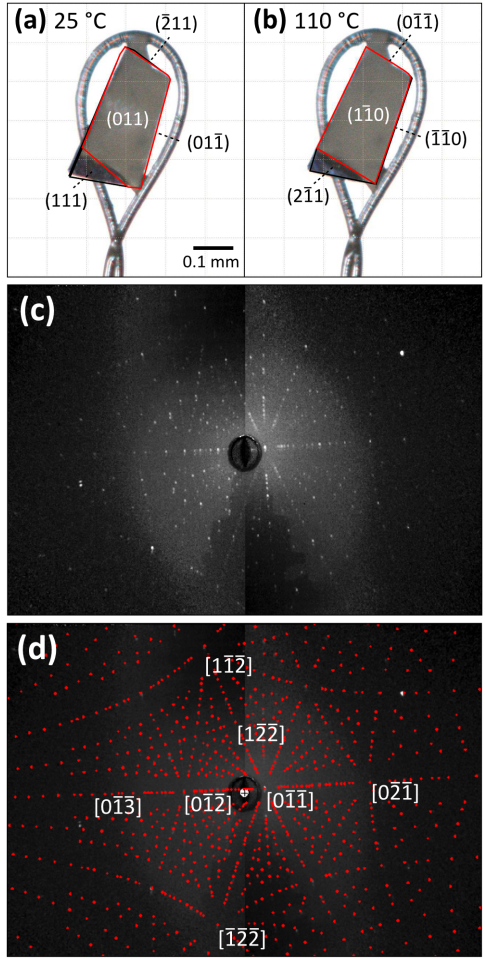

Figure 2: Determination of the surface orientation. (a, b) Crystal fragment investigated with XRD for temperatures below and above the MIT. The experimentally obtained Miller indices of several crystallographic planes are indicated. (c) Laue back-reflection measured at room temperature (monoclinic phase), rotated slightly out of normal incidence. (d) The same pattern superimposed by a fit.

*Surface techniques*: The surface properties of the $VO_2$ single crystals were investigated with XPS, LEED, and STM. The XPS measurements were conducted in an UHV chamber equipped with two hemispherical analyzers (Omicron 'ARGUS') in 80° geometry with overlapping detection areas, described elsewhere [33]. This setup allows for simultaneous acquisition of normal- and grazing-emission spectra using focused monochromatic Al Kα X-rays (1486.6 eV; Omicron XM 1000).

The STM and LEED experiments were performed in two different UHV systems: (1) A two-chamber UHV-system consisting of an analysis chamber (base pressure $1\times10^{-10}$ mbar) equipped with a variable-temperature STM (Aarhus 150, SPECS), a multi-channel plate (MCP) LEED (SpectaLEED, Omicron), and a preparation chamber (base pressure $<5\times10^{-10}$ mbar) for sample cleaning and preparation purposes. (2) A two-chamber UHV-system consisting of an analysis chamber (base pressure $2\times10^{-11}$ mbar) equipped with a low-temperature STM (Omicron LT-STM), and an adjacent preparation chamber (base pressure $5\times10^{-11}$ mbar) containing LEED and sample cleaning facilities. Both STMs used etched W tips and were heated to ~80°C.

*Surface measurement procedures*: For the measurements in UHV, the crystals were inserted into the chambers via a small side chamber without bake-out/heating, to minimize the number of phase transitions. It was found that each phase transition shortens the lifetime of the crystals significantly. However, even when minimizing the transitions to two, *i.e.*, cooling to room temperature after the crystal growth and the first heating in UHV (thereafter the crystal is kept above $T_C$), they eventually break along grain boundaries, most likely due to mechanical stress during transfers between different positions in the vacuum chamber and the forces applied by the mounting process. Thus, after the first heating above the MIT temperature, the $VO_2$ crystals were always held at 200–250 °C, *i.e.*, also during sputtering, LEED and XPS measurements, as well as overnight and at 80 °C inside the STM. In both the STM/LEED chambers, the surface of the $VO_2$ crystals was initially cleaned by several cycles of sputtering (1 keV $Ar^+$ ions, ~2.5 $\mu A/cm^2$, 10 min) and annealing in UHV (10 min). In the chamber with the variable-temperature STM, the annealing temperature was determined as >600 °C measured on the crystal and ~700 °C measured on the Ta sample plate using a pyrometer with an emissivity set to 0.8. In the LT-STM system, the annealing temperature of ~600 °C was measured via a thermocouple. The surface was daily refreshed either by a full cleaning cycle or by annealing in-between the STM measurements. Variations in the sample preparation, which did not change the surface according to STM examination, included: Annealing at 600 °C in $2\times10^{-6}$ mbar $O_2$ combined with cooling in $O_2$ until the temperature decreased to 300 °C; annealing in $1\times10^{-8}$ mbar $H_2$ at 450 °C, and dosing 10 Langmuir (L, 1 L = $1.33\times10^{-6}$ mbar·s) of water into the STM at 80 °C. In the XPS chamber, the crystals were sputtered for 60 min (1 keV $Ar^+$ ions, ~8 $\mu A/cm^2$) followed by annealing at 350–700 °C for 10 min. The samples were sputtered prior to each annealing step to 'reset' the history of the surface.

**DFT Calculations:** All calculations were performed with the Vienna *ab initio* Simulation Package (VASP) [34]. The projector-augmented wave (PAW) [35] method was employed for treating core electrons. For oxygen 6 valence electrons ($2s^2 2p^4$) and for vanadium 13 valence electrons ($3s^2 3p^6 3d^4 4s^1$) were expanded in a plane-wave basis set with an energy cut-off set to 500 eV. As the present study is focused on the metallic rutile phase, we chose a density functional theory-based description using the meta-GGA SCAN functional [36]. This functional was reported [37] as the best compromise between computational cost and accuracy in terms of lattice parameters and relative phase stability for the rutile and monoclinic $VO_2$ phases. In the present paper, all calculations were performed assuming non-magnetic $VO_2$ systems. While a recent study suggested a better description of the surface energies [22] with spin-polarized calculations, we found that the influence on the current presented results is rather small. A detailed discussion of these findings will be presented in a forthcoming publication together with a critical assessment of the performance of various DFT functionals in this system [38].

The Brillouin zone was sampled with a Γ-centered Monkhorst-Pack grid [39], using 6×6×9 k-points for the bulk rutile phase. For surface calculations the k-points grid was adjusted to obtain a comparable sampling of the surface Brillouin zone. Ionic relaxations were stopped when all residual forces became smaller than $10^{-2}$ eV/Å. All slabs were calculated with lateral cell dimensions corresponding to the optimized bulk lattice constants with a separating vacuum layer kept at 15 Å.

While the surface energies of the (1×1) terminations were calculated from the linear interpolation of total slab energies ranging from five to eight layers, the surface free energies of the off-stoichiometric (2×2) surface terminations were calculated using five-layer slabs with symmetric top and bottom surfaces and the bulk energy derived from the (1×1) slabs. All simulated STM images were generated in the Tersoff-Hamann approximation [40] with a bias voltage of +2 eV. To obtain more accurate description of the charge density in these calculations, the energy cut-off was increased by 30% with respect to the other calculations.

## 3. Results and Discussion

### 3.1. Diffraction Results

Before discussing the diffraction results it is useful to recall the crystallographic relationship between the monoclinic and tetragonal phases. The lattice parameters of both phases are summarized in Table 1. The transition of a crystallographic plane characterized by $(hkl)_T$ to a monoclinic (M) plane is described in ref. [41] by

$$\begin{pmatrix} h \\ k \\ l \end{pmatrix}_M = \begin{pmatrix} 0 & 0 & -2 \\ -1 & 0 & 0 \\ 0 & 1 & 1 \end{pmatrix} \begin{pmatrix} h \\ k \\ l \end{pmatrix}_T$$

Thus, for example, $(0-1-1)_M$ transforms into $(1-10)_T \equiv (110)_T$, $(-2-12)_M$ into $(111)_T$, and $(-202)_M$ into $(011)_T$.

| crystal system | space group | $T$ (K) | $a$ (Å) | $b$ (Å) | $c$ (Å) | α, γ (°) | β (°) |
|---|---|---|---|---|---|---|---|
| monoclinic (M1) | P2$_1$/C | 300 | 5.75 | 4.53 | 5.38 | 90 | 122.6 |
| tetragonal (T) | P4$_2$/mnm | 360 | 4.55 | 4.55 | 2.85 | 90 | 90.0 |

Table 1: Structural parameters of the VO$_2$ phases below and above the transition temperature $T_C$ = 67 °C. [41,42]

The orientation of the shiny side of the crystals was determined by XRD (both, the monoclinic and tetragonal phase), Laue back-reflection (RT, monoclinic phase) as well as LEED (200 °C, tetragonal phase). In XRD, see Figure 2(a,b), the face marked by a red line corresponds to the $(011)_M$ plane of the monoclinic form at 25 °C and the $(1-10)_T$ plane of the tetragonal modification at 110 °C, respectively. For all crystals, the best match for the obtained Laue diffraction pattern is the $(0-1-1)_M$ surface, an example is given in Figures 2(c,d), which corresponds to $(110)_T$ above the phase transition temperature.

In UHV, the first LEED experiments with a conventional LEED optics showed a rectangular pattern above $T_C$ that vanished very quickly, although no typical charging effects were observed. In the MCP LEED (which uses lower electron currents, ~2.6 µA/cm$^2$) the pattern was stable for at least half an hour. The LEED images of Figures 3(a,b), obtained after annealing the sample at 600 °C, show a (2×2) superstructure. This superstructure is discussed in more detail in the STM section. Less intense satellites next to the main diffraction spots in Figure 3(a) originate from the differently tilted macroscopic crystallites that were present in most samples. Some diffraction spots associated with the reconstruction vanish at certain energies (white circles in Fig. 3(b)). The ratio of the surface lattice parameters ($a_S$, $b_S$) taken from the LEED images $a_S^*/b_S^*$ = ~2.3 matches the expected value of the $(110)_T$ surface, *i.e.*, $b_S/a_S = \sqrt{2} \cdot a_T/c_T$ = 6.43/2.85 = 2.26 (see also Table 1).

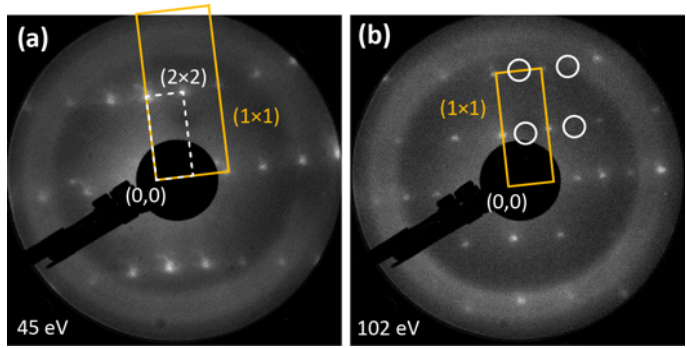

Figure 3: LEED pattern of the rutile phase at 200 °C taken at two different electron energies. The (1×1) pattern (solid orange rectangle) of a $(110)_T$ termination, and the (2×2) superstructure (dashed white rectangle) are indicated. Note that some reflections of the superstructure are not present in the image taken at 102 eV (white circles). Weak satellite spots are due to mosaicity.

**3.2 Scanning Tunneling Microscopy: Results and Analysis**

The surface structure observed in STM evolved with the annealing temperature after the sputter treatment. While the (1×1) structure is observed after mild annealing at ~560 °C, different superstructures emerge at higher temperatures. Annealing the crystal at ~560 °C (measured with a pyrometer on the crystal; 650 °C on the Ta plate) leads to a rather rough surface with terraces of only 5–10 nm size. On the terraces, the structure of rutile VO$_2$(110) is observed. This is shown in Fig. 4, where the atomic rows of the (110)$_T$ surface running in [100] direction are clearly visible, in particular in the right image. The absence of mirror symmetry in the island shape and the high step density suggest screw dislocations due to an STM tip crash nearby.

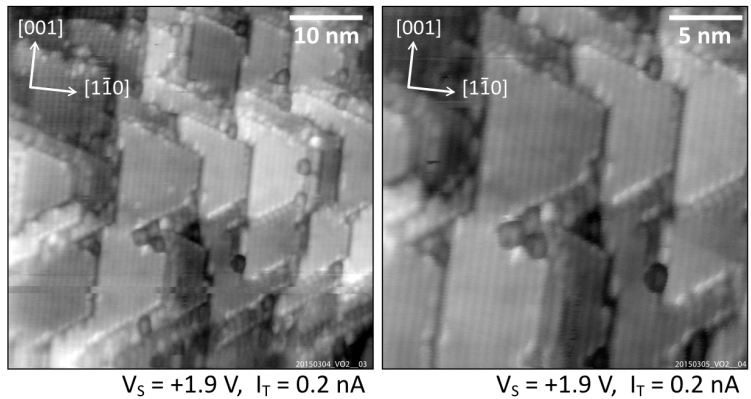

Figure 4: The VO$_2$(110)$_T$ surface after annealing at 560 °C, showing only the (1×1) surface without any superstructures in empty-states STM.

After annealing at >600 °C (as measured with the pyrometer on the VO$_2$ crystal), the surface exhibits terraces that are larger than after annealing at 560 °C, see Figure 5. The single-layer step height of VO$_2$(110)$_T$, ~320 pm (*i.e.*, $a/\sqrt{2}$, Table 1), was used to calibrate the z-distances. (This was necessary as the calibration of the piezo characteristics is usually done at room temperature and can be different at the elevated temperatures used in these STM measurements.) In addition to the unreconstructed VO$_2$(110)$_T$-(1×1) surface, two superstructures are found: A (2×2) and a c(4×2) overlayer; the latter was always on top of the (2×2) structure. Both were present on all crystals after annealing at >600 °C, but their coverages varied, see Figures 5(a, c). In the following, the (2×2) superstructure is described in detail to capture all features relevant for DFT modeling.

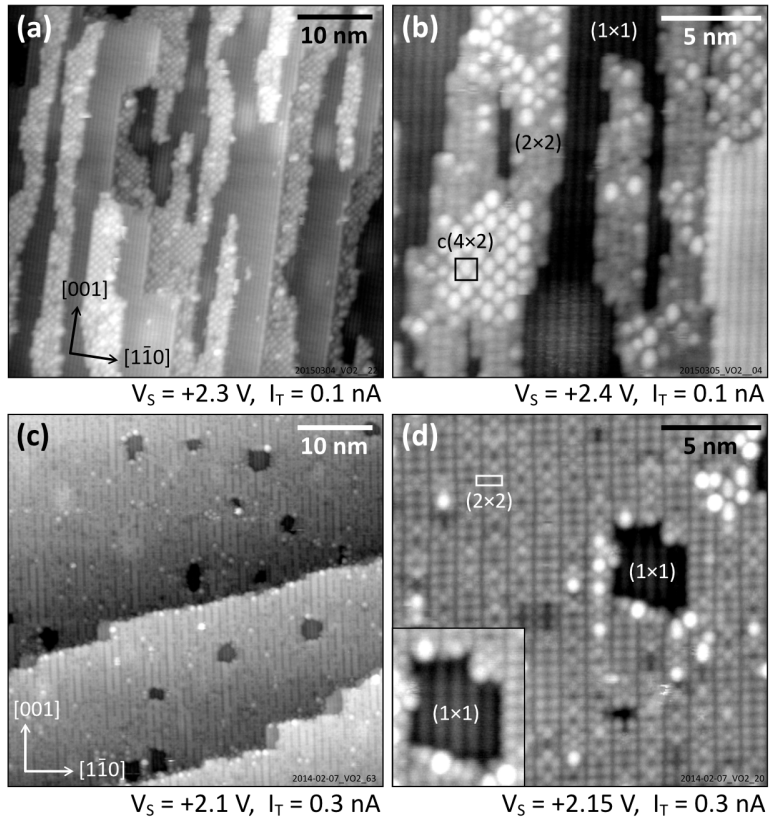

Figure 5: STM measurements on a VO$_2$(110)$_T$ crystal kept at 80 °C. (a) Overview image on a crystal with step edges along the [001] direction and a ~50% coverage by (2×2) and (1×1). (b) Detailed image showing the unreconstructed VO$_2$(110)$_T$- (1×1), narrow stripes of (2×2) and small patches of c(4×2). (c) Overview image on a different crystal with different step directions and an almost full overlayer with the (2×2) phase. (d) Detailed image showing a hole in the (2×2) layer.

The main superstructure/reconstruction is an adlayer with (2×2) symmetry in agreement with the LEED pattern. Figures 5(a, c) show overview images of the surface of two different crystals where the coverage of the (2×2) layer varies from ~50 % to almost 100 %. This can be explained by differences in the crystal temperature during annealing, caused by the uneven thermal contact with the sample plate due to the often irregular and rough backsides of the crystals or by different step densities and step orientations (miscut). In most preparations, the larger part of the surface was covered by the (2×2) structure, and the (1×1) VO$_2$(110)$_T$ surface was found only at step edges or in small holes of the adlayer, see Figure 5. The apparent height of the (2×2) structure on top of the VO$_2$(110) terrace measures ~160 pm. The (2×2) adlayer always grows attached to the lower step edges, regardless of the step orientation; usually it does not cover the terrace completely. On first glance the (2×2) structure consists of double rows parallel to the [001] direction, *i.e.*, two narrow-spaced rows separated by a wider spacing. Figure 6 provides STM images of this structure displaying four different contrasts. These have been taken at similar bias voltages and tunneling conditions; the change in contrast is purely tip-related. (The appearances displayed in Figs. 6(a) and (b) were observed most often.) The spaces between the double rows along the [001] direction are usually decorated with three different features, which do not depend on the imaging contrast of the (2×2) structure. They are all imaged as single protrusions sitting in the 'wide' sites, *i.e.*, between the pairs of narrow-spaced rows, seemingly in a 4-fold hollow site with respect to the protrusions of the (2×2) structure. The first species is an additional dot between the double-rows with a similar apparent height as the row maxima. These features often occupy every other '4-fold hollow site' along the rows, leading to flower-like features (white ellipses in Figure 6(a-c)). The second feature is a protrusion in the same site, but with a fuzzy appearance (indicated by white arrows in Fig. 6(c)) including 'fuzziness' of the four neighboring protrusions of the (2×2) structure. Both species are stable during image acquisition and do not diffuse at 80 °C. Finally, there are also a few very bright protrusions close to the sites that can be taken by the weaker maxima discussed here.

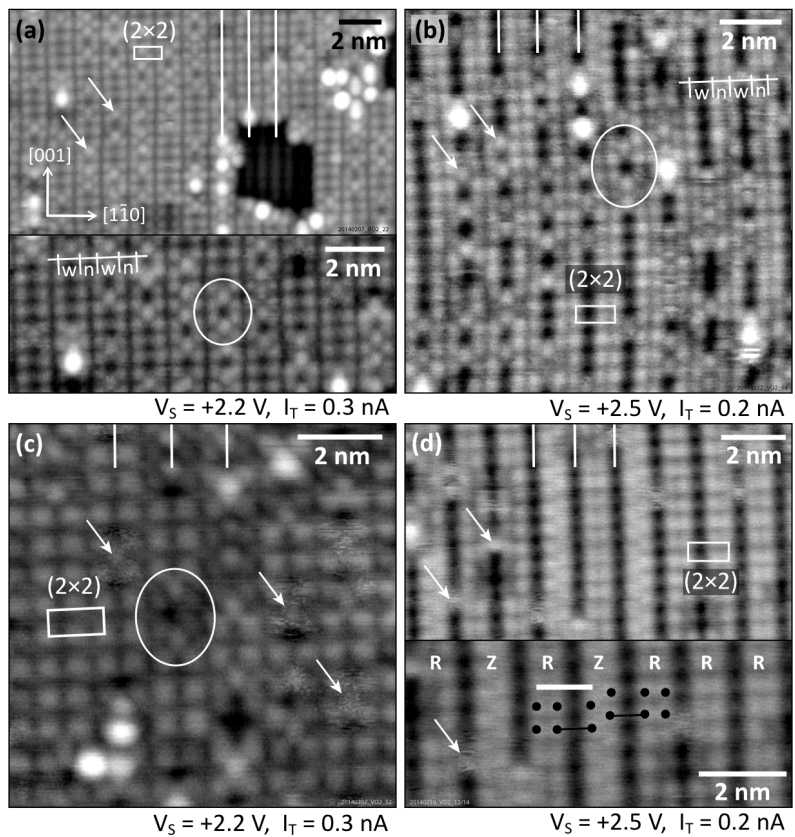

Figure 6: STM contrasts and details of the (2×2) row structure. White markers in the top part of each panel indicate the positions of the wide spacing between the double rows. Comparison of (a) and (b) shows that contrast does not always emphasize the wide spacing as the main depression. Additional features situated in the wide sites are marked by white arrows. (a, b) Most common appearances showing double rows. The alignment of the wide spacing with respect to the bright rows of the VO$_2$ surface visible inside the holes is indicated in (a). (c) Square appearance, *i.e.*, equal spacings in [1−10] are observed. (d) Distinct double row structure with domain boundaries (displacements within the double row along [100]), resulting in lines with zigzag (Z) structure, in contrast to the usual rectangular (R) arrangement.

All these features can be used to identify the wide spacing of the (2×2) double rows even if the rows appear equidistant as in Fig. 6(c). White lines at the top parts of Figure 6 mark the wide spacing – the narrow and wide spacings are indicated in panels (a,b), labeled 'n' and 'w', respectively. Measuring the distances of the individual protrusions within the ×2 periodicity along the [1−10] direction reveals different spacings for all contrasts presented in Figure 6. The structures in panels (a, b, d) show double rows, while the protrusions appear evenly spaced along [1−10] in panel (c). The structure in panel (d) is similar to (b) but it contains domain boundaries. The inset shows that these domain boundaries feature a zigzag (Z) arrangement of protrusions along the rows, instead of the expected rectangular (R) pattern. It should be noted that the two rows framing the wide spacing are always aligned in the [1−10] direction (indicated by connected dots in the inset in Figure 6(d)) and the zigzag is always within the double row.

The consistent alignment of the (2×2) double rows with respect to the bright rows of the VO$_2$(110)$_T$-(1×1) surface (see below), together with the identical features visible between the row pairs in all contrasts strongly indicate that all the structures presented in Figure 6 are, in fact, identical. The zigzag rows are the result of defective regions such as shifted building blocks or domain boundaries. The alignment of the (2×2) with respect to the (1×1) structure is evaluated in Figure 7.

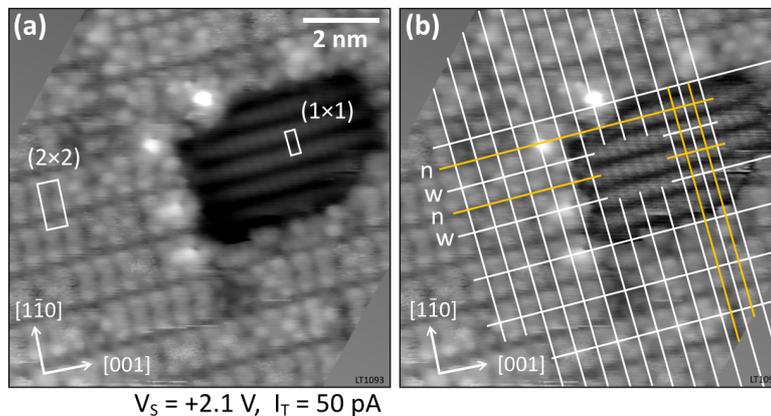

$V_S$ = +2.1 V, $I_T$ = 50 pA

Figure 7: Alignment of the (2×2) structure with respect to the VO$_2$(110)$_T$ (1×1). The grids in (b) pass on the (1×1) rows, but between the maxima of these rows (*i.e.*, they are offset by ½[001] with respect to the maxima of the (1×1) structure). Further explanations are in the text. The STM image of panel (b) is the same as in (a) with a high-pass filter applied to the (1×1) region.

Along the [001] direction, both the narrow and the wide spacings of the (2×2) structure are centered on the bright rows of the VO$_2$(110)-(1×1) area, *i.e.*, the protrusions of the (2×2) double rows are located in-between the rows of the (1×1) structure (see the lines in panel (b)). Inside the hole, where the VO$_2$(110)-(1×1) is visible, the lines of the (2×2) grid (white) in [1−10] direction are in-between the protrusions of the VO$_2$-(1×1) rows. To guide eye, the white and yellow lines together form the (1×1) structure. Along [1−10], the protrusions of the (2×2) structure are centered on every second line (yellow), hence they are in-between the protrusions observed in the VO$_2$(110) area. The second superstructure is often less prominent than the (2×2) and was not observed in the LEED pattern. It is a centered rectangular arrangement formed by bright, slightly elongated protrusions with lattice constants of ~1.29 and ~1.14 nm, *i.e.*, a c(4×2) periodicity with respect to VO$_2$(110) (1×1), or a c(1×2) with respect to the (2×2). It is always on top of the (2×2) layer and never on the VO$_2$ patches, although it is preferentially found in the vicinity of holes in the (2×2) layer (Figure 5(d)). Figures 5(a, b) show a very high coverage of this structure, where almost all of the (2×2) is covered. This structure is tentatively assigned to Cs impurities segregating to the surface, which were found in the XPS spectra discussed below. The Cs content differs from crystal to crystal, and it was not possible to change or remove it by excessive sputtering and annealing cycles including sputtering at 600 °C.

### 3.3. X-ray Photoelectron Spectroscopy: Chemical Composition

The surface and near-surface region of larger crystals was systematically investigated at different annealing temperature after cleaning by sputtering. The aim was to see whether the (2×2) and the c(4×2) structures are due to a stoichiometrically different vanadium-oxide phase or possibly related to the segregation of impurities. The investigated temperature range was 250–650 °C, and the measurements were repeated on 5 crystals. The main impurity seen on these samples was Cs, segregating at the surface above 450 °C (Cs 3d$_{3/2}$ and Cs 3d$_{5/2}$ peaks at 738.5 eV and 724.5 eV, respectively). After annealing at 650° C, the Cs concentration (calculated for Cs 3d$_{5/2}$ versus O 1s and V 2p, using atomic sensitivity factors) ranged from 0.1–2.3 % in the normal emission XPS spectra and 0.3–6.0 % in the respective grazing emission data and stayed constant for each crystal during the experiments. The XPS data shown in Figure 8 was obtained on a crystal with <1% Cs both in normal and grazing emission. Moreover, SiO$_2$ (Si 2p at ~103 eV and an additional O 1s shoulder at ~533.5 eV), and, occasionally, small amounts of Ni (from the thermocouple wire holding the crystals), Ar (incorporated during sputtering, normal emission only), Mo (sample holder, in grazing emission only), and K (with a concentration of 0.2–0.6 % in normal emission and 0.7–2.0 % in grazing emission; calculated for K 2p versus O 1s and V 2p using atomic sensitivity factors) were observed. The SiO$_2$ originates from the quartz crucible the VO$_2$ crystals were grown in. Possibly, some crystals grew directly attached to the crucible wall, *i.e.*, the SiO$_2$ is predominantly at the edges or backside of the crystal, which were also visibly discolored on some samples. Si was not detected in the LA-ICP-MS (trace analysis) across several other samples, but these samples (some of them had been previously used for STM) were selected to be visually without fault.

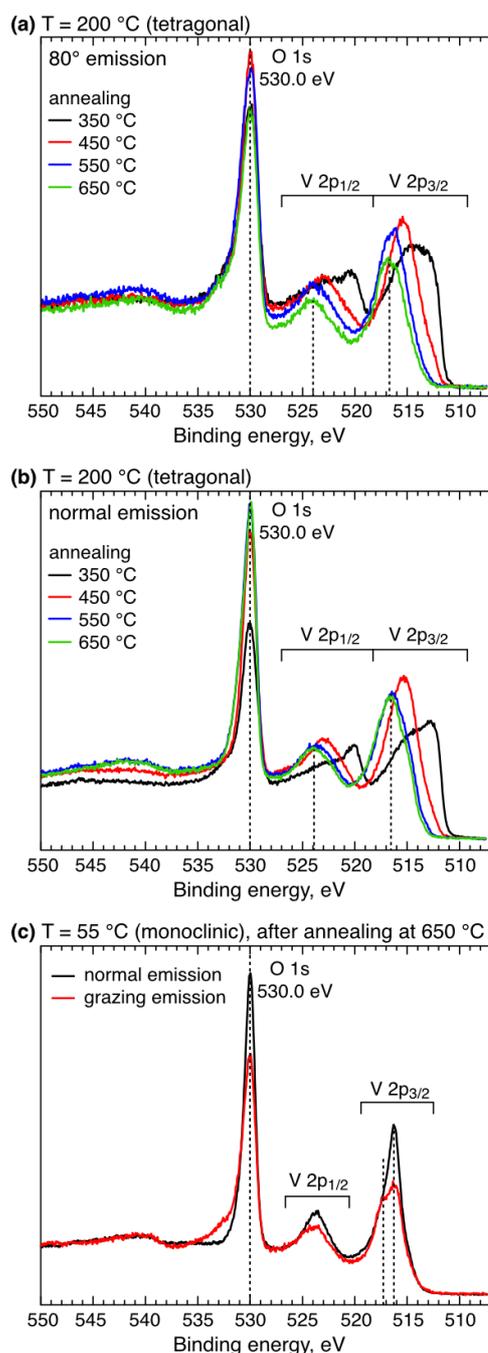

Figure 8: Temperature-dependent XPS core-level spectroscopy of a VO$_2$ single crystal. (a, b) The VO$_2$(110)$_T$ phase after annealing in UHV at different temperatures. The surface was sputtered before each annealing step. All spectra were acquired at 200 °C in (a) grazing emission and (b) normal emission. (c) Normal and grazing emission spectra taken at 55 °C (*i.e.*, from VO$_2$(011)$_M$), after annealing at 650 °C (green curves of (a, b)).

The evolution of the vanadium and oxygen core levels was followed with XPS as a function of the annealing temperature from 350 °C to 700 °C by simultaneously acquiring grazing emission (80°) and normal-emission (0°) spectra. Figure 8 displays the evolution of the O 1s and V 2p states measured at 200 °C, *i.e.*, tetragonal VO$_2$(110). The spectra are normalized with respect to the background on the low binding-energy side, and the binding energy of the O 1s peak is set to 530.0 eV for better comparison (compensating a gradual shift of total ~1.5 eV to lower binding energies; see Supplemental Material). The V 2p core level spectra are complex in both, grazing and normal emission (Figure 8(a) and (b), respectively), suggesting the presence of several oxidation states. Overall, the O 1s intensity increases relative to the V 2p intensity with increasing temperature. Quantitative fitting of these spectra is beyond the scope of this work, as one would have to take into account possible oxidation-state dependent satellite features of the V 2p levels, which overlap with the O 1s region and thus strongly affect the deconvolution into various components. Nevertheless, qualitative information can be deduced, and allow for comparing these data with results from the literature. Reported V $2p_{3/2}$ binding energies for different oxidation states vary from ~512.35 eV for V(0) to ~517.20 eV for V(V) (with the O 1s located at 530 eV) [43,44]. With increasing annealing temperature, the V 2p signals in Figure 8 shift to higher binding energies. At 350 °C (black spectra), where the surface is still roughened from the ion bombardment, oxidation states range from metallic V(0) to V(V), with an emphasis on metallic and/or low oxidation states. At 450 °C (red) the V $2p_{3/2}$ feature centers around V(III) (~515.3 eV), with a shoulder towards the metallic side. At temperatures of 550 °C (green) and higher (650 °C, blue), the V $2p_{3/2}$ binding energy lies between those of V(IV) and V(V) (~515.8 eV and ~517.2 eV, respectively [43]), indicating a mixture of these oxidation states. The comparison of normal and grazing emission data reveals a small shift (0.1–0.2 eV) towards higher binding energies at 80°, indicating subtle changes towards higher oxidation states at the surface.

In Figure 8(c), the monoclinic phase was investigated with XPS [45] by preparing the sample at 650 °C and cooling below the MIT temperature to 55 °C. Spectra obtained by this procedure essentially display the same oxidation states as in the tetragonal phase but with a drastically different peak shape due to the different screening in the metallic and insulating phases [46]. A plot comparing the tetragonal and monoclinic spectra is provided in the Supplemental Material (Figure S4).

### 3.4. DFT: The $(110)_T$ (1×1) and $(011)_T$ (1×1) terminations of rutile $VO_2$

The slab calculations were performed at the optimized bulk-like lattice constants presented in Table 2. In the case of the rutile bulk, both the c/a ratio and the volume of the cell are underestimated by about 3 %. The SCAN functional correctly predicts the rutile phase to be metallic. The calculated lattice vector *c* is related to the vanadium-vanadium pairing and is in perfect agreement with previous reports for the same functional [37].

|  | a (Å) | c (Å) | c/a | V (Å$^3$/f.u.) | $\sigma_{110}$ (meV/Å$^2$) | $\sigma_{011}$ (meV/Å$^2$) |
|---|---|---|---|---|---|---|
| SCAN (this work) | 4.56 | 2.77 | 0.61 | 28.81 | 40 | 76 |
| Expt. [41,42] | 4.56 | 2.86 | 0.63 | 29.63 | - | - |

Table 2: Calculated parameters of the rutile unit cell and surface energies σ of $VO_2$, using the meta-GGA (SCAN) functional.

The stability of the unreconstructed low-index facets of the rutile phase, $(110)_T$ and $(011)_T$ was evaluated by calculating their respective surface energies (Table 2) finding the (110) surface to be more stable by 36 meV/Å$^2$.

The bulk-terminated, relaxed surfaces of rutile $VO_2(110)_T$ and $VO_2(011)_T$ are shown in Figures 9(a,b), respectively. In calculated STM images of the $VO_2(110)_T$ terminated surface, Figure 9(c), the twofold-coordinated oxygen atoms appear as straight, bright chains with a distance of 6.45 Å. Note that this appearance is different from that of the well-known $TiO_2(110)$ surface, where the bridging oxygen rows appear dark, [47] but the same as that of $RuO_2(110)$, which is also metallic [48]. Both, the appearance of rows and their separation is in agreement with experimental STM images of $VO_2(110)_T$ (1×1) in Figures 4 and 5(a). The bright features on the $VO_2(011)_T$ termination, Figure 9(d), are formed by both vanadium and oxygen atoms, resulting in zigzag chains.

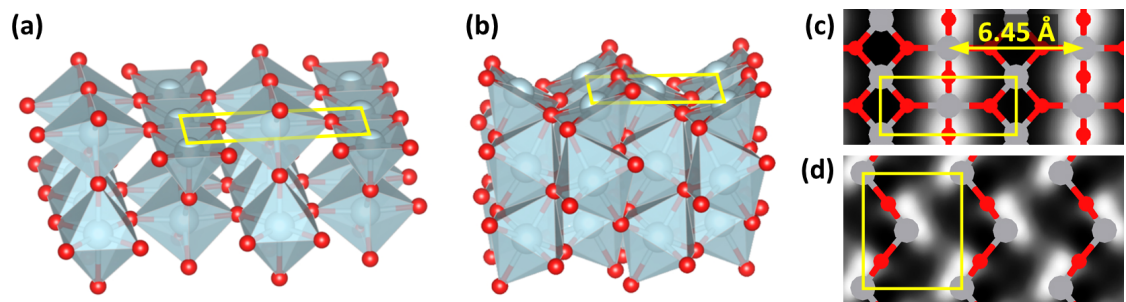

Figure 9: Side views and simulated STM images (empty states) of (a, c) the $VO_2(110)_T$ and (b, d) the $VO_2(011)_T$ bulk terminations using the Tersoff-Hamann approximation.

### 3.5. Developing a Surface Model

Evidently, the experimentally observed (2×2) phase cannot be obtained by a simple variation of the bulk structure due to the lower symmetry. To investigate possible reconstructions of the bare $VO_2(110)_T$ surface, a simulated annealing technique was employed. A subsequent relaxation (with the SCAN functional) shows that even for the bare $VO_2(110)_T$ surface, the calculated ground state has its symmetry lowered due to a buckling in the topmost layer, resulting in a (2×1) superstructure (Fig. 10, shown together with a simulated STM image). The surface reconstruction is formed by the displacement of vanadium atoms along the [1−10] direction, leading to a relative height difference of 0.31 Å between the vanadium atoms. Nevertheless, with an energy gain of only 80 meV per (2×1) unit cell (SCAN) the buckled ground state is almost degenerate with the unreconstructed surface termination. As a side note, other GGA and meta-GGA functionals also predict a similar trend: using PBE, PBE+U (U-J = 2 eV) and SCAN+rVV10 functionals; the buckling stabilizes the surface by 75, 39 and 34 meV, respectively.

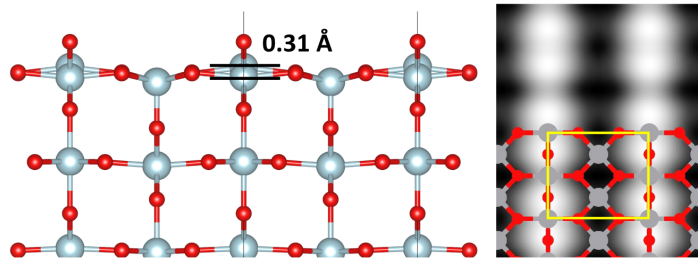

Figure 10: Buckled superstructure on VO$_2$(110)$_T$ and the corresponding calculated STM image. The structure was obtained by simulated annealing and relaxed using the SCAN functional. Bright spots in the simulated STM image are formed by the topmost oxygen row. Note that all oxygen atoms have the same height.

A molecular dynamics simulation performed at a temperature of 350 K with four-layered slabs to reduce the computational effort showed that the vanadium pairs can flip from the up-down to down-up configuration and vice versa with the average flipping time of 200 fs. Since STM measurements show time-averaged data and the flipping time is far below the resolution limit, this buckled superstructure will exhibit the same STM image as an unbuckled VO$_2$(110)$_T$ surface.

Another hint guiding one towards an appropriate atomic model for (2×2) structure can be derived from the experimental findings, which suggest a surface stoichiometry different from the bulk: Our XPS measurements indicate an increased concentration of O on the surface and the presence of V$^{5+}$ cations. In addition, the experimental STM images of the (2×2) resemble patterns found on the V$_2$O$_5$(001) surface as characterized by Blum et al. [31], suggesting that a vanadium pentoxide monolayer might be a good starting point for the development of an atomic model. We take the orientation of the V$_2$O$_5$ lattice such that the cleavage plane is (001), *i.e.*, the V=O vanadyl bonds are roughly parallel to [001]. To fit to the VO$_2$(110)$_T$ substrate in a (2×2) configuration, the unit cell of the V$_2$O$_5$(001) monolayer needs to be expanded in [100] direction from 11.50 Å to 12.86 Å (*i.e.*, by 12%), breaking up the structure along the dashed symmetry plane, and in [010] direction from 3.56 Å to 5.70 Å (*i.e.*, by 60%), see Figure 11(a). This strong distortion leads to the rearrangement of the V$_2$O$_5$ building blocks, namely a change from edge-sharing pyramids to corner-sharing tetrahedra. The dark vanadium polyhedra pointing away from the surface (towards the vacuum) shift along the [010] direction, and the inverted (bright) polyhedra are pulled towards each other in [100] direction. This is marked by yellow arrows in Figure 11(a). The result is a hexagonal ring of vanadium tetrahedra as shown in Figure 11(b). A similar structure consisting of corner-sharing up- and down-pointing VO$_4$ tetrahedra has already been confirmed for vanadium oxide on a Pd(111) surface [30]. For the unsupported model layer (Figure 11(b)), the lateral distance between the oxygen atoms at the top of the tetrahedra along [1−10] is 3.8 Å. When this layer is supported by the rutile (110) surface as shown in Figure 11(c) and (d), the distance is slightly decreased to 3.7 and 3.6 Å, respectively. In both cases, the surface has an overall stoichiometry of V$_4$O$_{13}$.

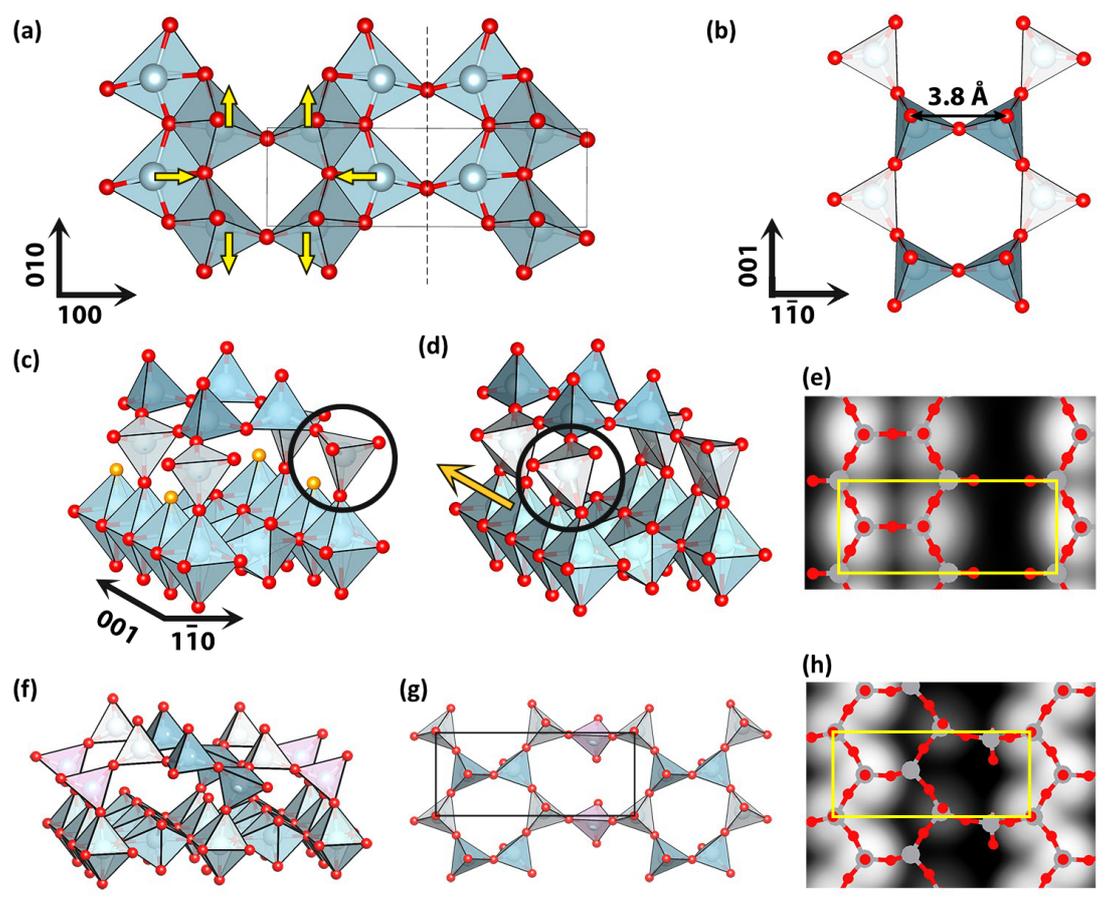

Figure 11: Conceptual steps towards the VO$_2$(110)$_T$ ring terminations. (a) A vanadium pentoxide (001) monolayer. The yellow arrows mark the displacement direction of the vanadium atoms, which causes the transformation to the tetrahedral 'ring superstructure' that fits the VO$_2$(110)$_T$-(2×2) supercell, shown in (b). Panels (c, d) show side views of two possibilities how this ring structure could connect to the underlying VO$_2$(110)$_T$ lattice, resulting in an overall stoichiometry of V$_4$O$_{13}$. The purely tetrahedral ring termination (panel c) contains in the subsurface layer V=O vanadyl bonds that can be subsequently removed, see oxygen atoms colored in orange. The difference between these structures is a shift along the [001] direction as it is pointed with the yellow arrow in panel (d), changing the coordination geometry of marked vanadium atoms from tetrahedra to square pyramids. The configuration in (d) is lower in energy. Figure (e) shows a calculated STM image of the ring structure in (d). Panels (f, g) show side and top views of another ring termination that proceeds from (b). The calculated STM image of this structure is shown in panel (h).

Figures 11(c) and 11(d) show two different configurations indicating how the structure could be placed on the $VO_2$ substrate underneath. In the first case, panel (c), the ring termination is bound in a corner-sharing fashion, *i.e.*, with just with one oxygen bond, forming a purely tetrahedral termination, and leaving half of the undercoordinated O atoms of the substrate unterminated. These atoms, colored in orange, can be also partly or fully removed, which leads to $V_4O_{12}$ and $V_4O_{11}$ surface stoichiometries. In Figure 11(d), the ring termination is shifted by a half of the (1×1) rutile [001] lattice vector as marked with the yellow arrow. The additional bond to the undercoordinated O atoms of the substrate layer converts the lower tetrahedra to square pyramids, see black circles in panels (c, d). This shift stabilizes the surface termination by 0.24 eV per (2×2) supercell. Figure 11(e) shows a simulated STM image of the ring structure from the Figure 11(d). The image consists of pairs of spots with a separation of 3.6 Å, in agreement with the experimental data for the narrow distances of the bright features from panel 6(b). The ring pattern can be transformed into another stable structure shown in Figs. 11(f, g) by two major changes. First, the rings are connected by another vanadium tetrahedron (depicted in violet color), which also changes the surface stoichiometry to $V_5O_{14}$. Second, half of the dark tetrahedra are binding to the substrate layer, while the other tetrahedra are moved to the surface layer (light gray in Figure 11(f,g)). This modified ring structure results in a zig-zag STM pattern shown in Figure 11(h). The bright spots are separated by 2.1 Å (projected in [1−10] direction), in good agreement with the experimental value of 2.7 Å for the zig-zag pattern in Figure 6(d). The simulated STM image in Fig. 11(f) shows an additional subtle spot that comes from the second row of upward-pointing tetrahedra. The height difference of the topmost oxygen atoms between the rows made of upward-pointing tetrahedra is approximately 1 Å, thus the low-lying protrusions are not expected to be observed in the experiment. All ring structures, Figs. 11 (c, d, f), are thicker (5.5, 5.4 and 5.8 Å, respectively) than a $VO_2(110)_T$ rutile layer with thickness of 3.3 Å.

To explain the additional features in the STM images, other metastable ring-type structures were also explored. In these structures, an additional vanadium tetrahedron was added between the rows, which changes the stoichiometry of the surface layer ($V_5O_{14}$ and $V_5O_{15}$); these structures are less stable than those in Fig. 11(c–h). Details are discussed in the Supplemental Material.

## 3.6. Stability of Surface Phases

To evaluate the stability of various surface terminations, we plotted the surface free energy as a function of the oxygen chemical potential as it is described in Ref. [49] – see Figure 12. The black, horizontal line represents the stoichiometric buckled $VO_2(110)_T$ surface. Green lines denote (2×2) supercells of this buckled surface with 1, 2 or 4 additional O atoms adsorbed in a vanadyl configuration on top of the fivefold coordinated V. For the latter two cases, our preferred structures agree with the models of an earlier DFT study by Mellan *et al.* [24]. Decreasing the coverage from 1/2 to 1/4 ML (1 adsorbed oxygen atom) every second oxygen atom is removed from the remaining oxygen row. Blue, orange, pink and red lines mark the oxygen-rich ring-superstructures, including $V_4O_{13}$, $V_5O_{14}$ (zig-zag and ring), and $V_5O_{15}$ stoichiometries as depicted in Figures 11(d,f), S5(a,b) respectively. Furthermore, gray dashed lines represent the reduced $V_4O_{13}$ ring structure depicted in Figure 11(c) where the undercoordinated oxygen atoms in the subsurface layer (colored in orange) were subsequently removed from both rows, leading to the $V_4O_{12}$ and $V_4O_{11}$ surface stoichiometries.

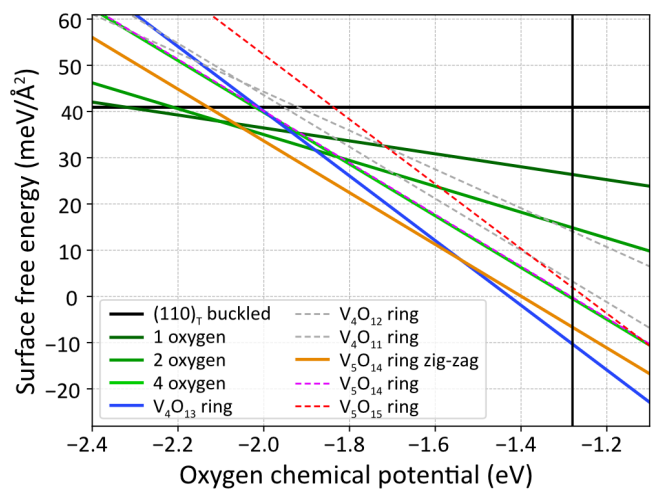

Figure 12: Calculated surface free energies (SCAN functional) of the $VO_2(110)_T$ surface as a function of the oxygen chemical potential. Shown are the buckled $VO_2(110)_T$ termination (from Fig. 10a, black line), oxygen adsorption on the buckled $(110)_T$ surface (green lines), and the ring terminations from Figures 11(d) (blue line), 11(f) (orange line) and S5(a,b) (dashed lines). Gray dashed lines represent the reduced $V_4O_{13}$ ring termination from Figure 11(c) obtained by a subsequent removal of oxygen atoms (colored in orange) from the subsurface layer that create V=O vanadyl bonds, leading to the $V_4O_{12}$ and $V_4O_{11}$ surface stoichiometry. The solid black, vertical line represents the experimental stability limit of the $VO_2$ phase with respect to the vanadium pentoxide phase.

Structures and simulated STM images of additional ring terminations are presented in the Supplemental Material. The plot also shows the stability limit of the $VO_2$ phase with respect to the vanadium pentoxide as a vertical black solid line defined as the enthalpy of the following reaction: $2VO_2 + ½ O_2 \rightarrow V_2O_5$. For calculating this phase boundary, the experimental heats of formation of the $VO_2$ and $V_2O_5$ phases with respect to vanadium metal [50,51] were used. It should be noted that the calculated phase boundary strongly depends on the chosen functional and spin-treatment, due to the peculiarities of an appropriate treatment of the $VO_2$ phase. These dependencies will be discussed in detail in a forthcoming publication [38].

Over a wide range of chemical potentials, the ring structure with $V_4O_{13}$ (Figure 11(d)) and the zig-zag with $V_5O_{14}$ (Figure 11(f,g)) surface stoichiometry are the most stable configurations. An unreconstructed, buckled $VO_2(110)_T$ surface, partially covered with O atoms would be stable under strongly reducing conditions (oxygen chemical potential less than −2.05 eV).

## 3.7 Discussion

This work clearly shows that the lowest-energy state of the $VO_2$ surface in a wide range of chemical potentials is a reconstruction, distinctly different from a bulk-terminated surface. The unreconstructed rutile $(110)_T$ termination is found only after mild annealing of a sputter-treated surface. After equilibration at higher temperatures, an adlayer with a double-row superstructure is observed. While the (2×2) periodicity is consistent with previous reports [21,22]; the simple models proposed earlier [24] that invoke only adsorption of excess O are neither supported by the STM measurements nor by DFT calculations.

The DFT models explain the main features of the STM images. The calculations showed that the aligned, bright spots (separated by 3.6 Å) in Figure 6(b) are related to the ring structure, which is the most stable surface termination at an oxygen chemical potential of −1.54 eV and higher. The model assigns the experimental double rows in Figs. 6 (a, b) to O atoms at the apex of $VO_4$ tetrahedra. In the range of chemical potentials between −1.54 and −2.05 eV, another ring structure is more stable that exhibits the zig-zag pattern similar to the features observed in Figure 6(d).

The stability of the ring terminations, especially at chemical potentials corresponding to higher oxygen pressures, is related to both the fact that the ring structure contains more oxygen than the adsorption phases and also to the close relationship of the ring structure to a vanadium pentoxide monolayer whose surface energy is only 11 meV/Å$^2$ according to our calculations. This relationship is not only structural – as we pointed out, the ring structures were derived from a $V_2O_5$ monolayer – but also evident in the electronic structure. As shown in Figure 13, the projected density of states (pDOS) onto vanadium and oxygen atomic orbitals of the ring phase compares well with the $V_2O_5$ bulk pDOS. The graphs show that, unlike the $VO_2$ phase, $V_2O_5$ as well as the $V_4O_{13}$ ring display a band gap where the V 3d band is separated by 1.9 and 2.2 eV from the O 2p band, respectively. It should be noted that the calculated band gap for the $V_2O_5$ phase underestimates the experimental band gap of 2.2–2.4 eV [52-54] and, therefore, we also expect a similar underestimation for the surface phase.

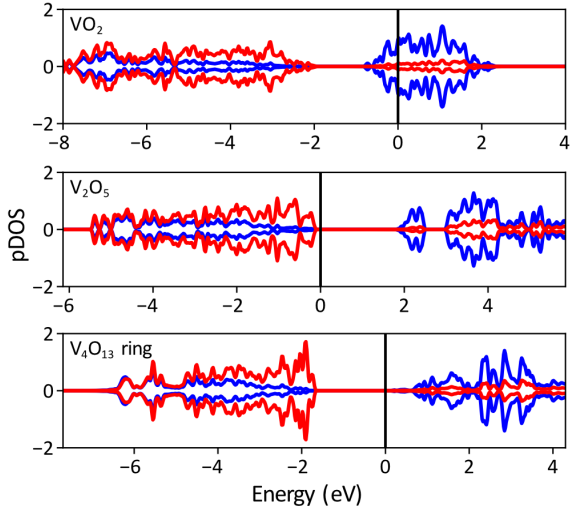

Figure 13: The (SCAN) projected density of states (pDOS) onto vanadium and oxygen atomic orbitals in the $VO_2$, $V_2O_5$ bulk phases, and in the $V_4O_{13}$ ring termination, normalized to one atom each and aligned with respect to the upper edge of the O 2p band. Oxygen and vanadium projections are colored in red and blue respectively. The black vertical line denotes the Fermi level which in the case of the ring termination is determined by the $VO_2$ substrate.

While we find a similar stability for the oxygen adsorption phases on the $VO_2(110)_T$ surface as discussed by by Mellan *et al.* [24], reconstructions were not considered in that computational study. Considering that a simple adsorption phase cannot explain the atomically resolved STM measurements in the present work, we conclude that the ring terminations are central building blocks for an atomistic understanding of the surface termination of $VO_2(110)_T$.

Comparing our favored model (Figure 11(d)) and STM simulations with the experimental STM images (Figure 7), we find that the alignment of the bright spots confirms that the holes in the experimental images revealing the bare (1×1) surface are a double layer below the ring surface structure. However, the experimental apparent height difference in STM (1.6 Å) is much smaller than the calculated geometrical thickness of the double layer (5.4 Å). This might be explained by the insulating nature of the $V_4O_{13}$ surface layer.

However, not all experimental features are captured in the present model. First, experimental STM images show the presence of additional bright spots between the double rows that are not explained by the $V_4O_{13}$ ring termination (ovals in Figs. 6). Second, the double rows in the experiment are always aligned with respect to the neighboring row like as in Figs 6(a,b), but our present model also allows the hexagonal rings that form the double-row pattern to be shifted by half of the unit cell in [001] direction, which is not observed in the experiment. The effect that would restrict the observed structure just to the aligned pattern is not evident from the DFT model; probably the alignment is caused by the entities forming the additional bright spots between the double rows. Furthermore, the experimental zig-zag row always neighbors a rectangular row (see Fig. 6d), which is not captured by our model because of the limited periodicity of the considered (2×2) supercell.

Nevertheless, the ring structures with the $V_4O_{13}$ and $V_5O_{14}$ surface stoichiometries can be seen as a structural basis that explains the most prominent features of the experimentally observed surface reconstruction. However, it should be pointed out that a huge amount of potentially stable surface structures exists, with only subtle energy differences between them. A further exploration remains a challenging task for the near future, see *e.g.* Ref. [38].

## 4. Summary

In summary, this work reports a comprehensive study on the surface properties of $VO_2$ single crystals, employing imaging, diffraction and spectroscopy techniques complemented by DFT calculations. The crystals exhibit the expected bulk structures below and above the MIT, and the most stable surface, assigned as (0−1−1) and (110) in the monoclinic and tetragonal structure, respectively, was investigated in detail. Impurities show strong variations even within individual samples, but are not affecting the described results (except for a minority superstructure with c(4×2) symmetry that is attributed to Cs atoms). XPS in grazing (surface sensitive) and normal (more bulk sensitive) emission shows that the surface is oxygen rich. The distinctly different XPS peak shapes reflect the drastic change in electronic structure that accompanies the MIT.

The most prominent outcome of the present work is a description of a (2×2) surface phase that is structurally distinct from the $VO_2$ bulk. The proposed ring models are based on corner-sharing tetrahedra and pyramids structurally and electronically similar to $V_2O_5$ layers, implying that surface atoms are in a markedly different environment than the bulk atoms, likely with profound influence on the surface properties, such as oxygen adsorption or the temperature of the metal-insulator transition of nanoparticles of this interesting material.

**Acknowledgements**

Support by the Austrian Science Fund FWF, projects V 773-N (Elise-Richter-Stelle) and Z 250-N27 (Wittgenstein Prize), is gratefully acknowledged. The authors further thank Andrey Prokofiev and Flora Pötzleitner for Laue backscattering and initial XPS measurements, respectively.

The samples were synthesized by L.A.B. STM and LEED measurements were performed by M.W. XPS measurements were done by M.W. and B.S.J.H. Trace analysis measurements were done by A.L. XRD was applied by J.L. The experimental data was analyzed by M.W., B.S.J.H., F.M., H.-P-.S., and M.S. DFT calculations were performed by J.P and F.M., and analyzed by J.P., F.Mi., M.S., and J.R. All authors discussed the data and contributed to the paper. M.W., J.P., F.Mi., and U.D. wrote the paper. U.D. and F.Mi. oversaw the project.

# Supplemental Material

**An oxygen-rich, tetrahedral surface phase on high-temperature rutile VO$_2$(110)$_T$ single crystals**

Margareta Wagner[1*], Jakub Planer[1,2], Bettina S. J. Heller[3], Jens Langer[4], Andreas Limbeck[5], Lynn A. Boatner[6], Hans-Peter Steinrück[3], Josef Redinger[1,2], Florian Maier[3], Florian Mittendorfer[1,2], Michael Schmid[1] and Ulrike Diebold[1]

[1] *Institute of Applied Physics, TU Wien, Wiedner Hauptstraße 8–10/134, 1040 Vienna, Austria*
[2] *Center for Computational Materials Science, TU Wien, Wiedner Hauptstraße 8–10/134, A-1040 Vienna, Austria*
[3] *Chair of Physical Chemistry II, Friedrich-Alexander-Universität Erlangen-Nürnberg (FAU), Egerlandstrasse 3, 91058, Erlangen, Germany*
[4] *Chair of Inorganic and Organometallic Chemistry, Friedrich-Alexander-Universität Erlangen-Nürnberg (FAU), Egerlandstrasse 1, 91058 Erlangen, Germany*
[5] *Institute of Chemical Technologies and Analytics, TU Wien, Getreidemarkt 9/164, 1060 Vienna, Austria*
[6] *Materials Science and Technology Division, Oak Ridge National Laboratory, Oak Ridge, Tennessee 37831, USA*

## 1. Experimental

### 1.1. Bulk characterization: Trace Analysis by LA-ICP-MS

The quantification was done using NIST612 with $^{51}$V as an internal standard. Table S1 contains the averaged value measured across five samples. Some elements are very inhomogeneously distributed in specific samples; see also the standard deviation given in Table 2. Figs. S1-3 are examples of element distributions across three samples, always showing the constant signal of $^{51}$V in comparison. The *x*-axis (time) corresponds to a line scan across the width of the crystals, which was ~1-2 mm in all cases.

While most impurities are of unknown origin, the high amounts of Ta and Au in sample A are due to the sample treatments in UHV. Sample A had been mounted on a Ta sample plate using a Ta small clip on the sample and Au foil underneath the crystal for better thermal contact. The crystal was cleaned in UHV by sputtering and annealing. Thereby, Ta was sputtered on the crystal surface, and Au was distributed either via sputtering or by diffusion during annealing at high temperature.

---

[*] Corresponding author: wagner@iap.tuwien.ac.at

|    | A_1   | A_2   | B_1   | B_2   | C_1   | C_2   | D_1   | D_2   | E_1   | E_2   |
|----|-------|-------|-------|-------|-------|-------|-------|-------|-------|-------|
| Ba | 0.17  | 0.03  | 0.28  | 0.13  | 2.21  | 0.83  | 1.15  | 2.09  | 0.32  | 0.24  |
| Ta | 301.69| 240.73| 0.47  | 0.72  | 0.16  | 1.09  | 0.03  | 0.03  | 14.30 | 5.71  |
| Au | 57.26 | 26.47 | 11.92 | 14.11 | 0.01  | 0.00  | 0.09  | 0.14  | -0.02 | -0.01 |
| Pb | 0.35  | 0.10  | 0.66  | 0.91  | 6.36  | 2.50  | 0.99  | 1.58  | 0.11  | 0.93  |
| U  | 6.97  | 4.31  | 8.85  | 9.78  | 29.02 | 17.53 | 26.04 | 40.79 | 0.25  | 0.23  |
| Mg | -0.01 | 0.00  | 0.00  | 0.00  | 0.00  | 0.29  | 0.01  | -0.01 | 0.41  | 0.41  |
| Al | 0.03  | 0.03  | 0.08  | 0.03  | 0.04  | 0.05  | 0.09  | 0.08  | 80.47 | 78.84 |

Table S1: Elemental distribution of five samples (A-E), with two lines across each sample. All values are given in µg/g (ppm).

|    | A_1    | A_2    | B_1   | B_2   | C_1   | C_2   | D_1   | D_2   | E_1   | E_2   |
|----|--------|--------|-------|-------|-------|-------|-------|-------|-------|-------|
| Ba | 0.44   | 0.10   | 0.39  | 0.07  | 1.57  | 0.75  | 0.67  | 0.83  | 0.09  | 0.08  |
| Ta | 683.14 | 474.65 | 0.38  | 0.12  | 0.05  | 1.40  | 0.02  | 0.02  | 24.80 | 0.89  |
| Au | 73.75  | 35.81  | 29.05 | 21.56 | 0.05  | 0.04  | 0.05  | 0.04  | 0.03  | 0.02  |
| Pb | 1.01   | 0.32   | 0.91  | 0.84  | 4.29  | 3.21  | 0.53  | 0.71  | 0.12  | 2.21  |
| U  | 4.92   | 2.37   | 6.68  | 1.97  | 13.93 | 7.51  | 8.33  | 14.00 | 0.06  | 0.15  |
| Mg | 0.05   | 0.01   | 0.00  | 0.00  | 0.01  | 1.31  | 0.01  | 0.01  | 0.07  | 0.11  |
| Al | 0.02   | 0.01   | 0.15  | 0.00  | 0.01  | 0.06  | 0.08  | 0.02  | 4.31  | 4.55  |

Table S2: Standard deviation of the trace measurements. Values in red indicate a very inhomogeneous distribution of the respective element in the sample.

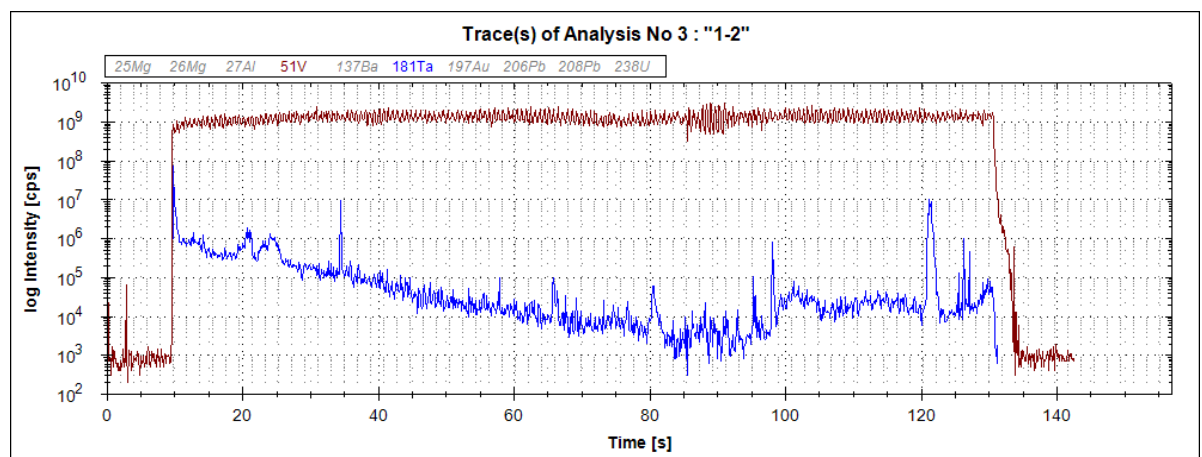

Figure S1: Sample A, line 1: The $^{181}$Ta intensity is variable across the sample, while the $^{51}$V signal remains constant.

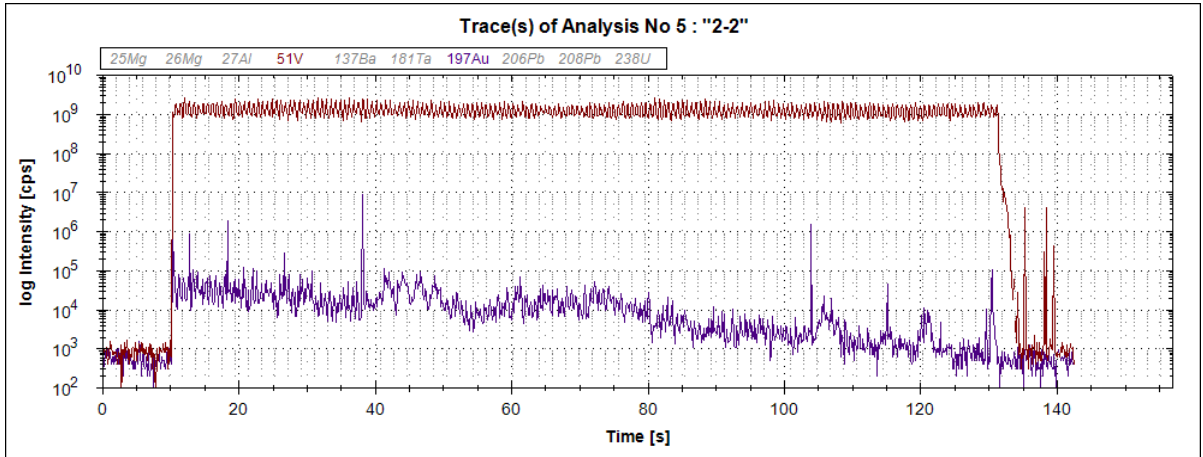

Figure S2: Sample C, line 1: The inhomogeneous intensity of $^{197}$Au in comparison to the constant $^{51}$V signal.

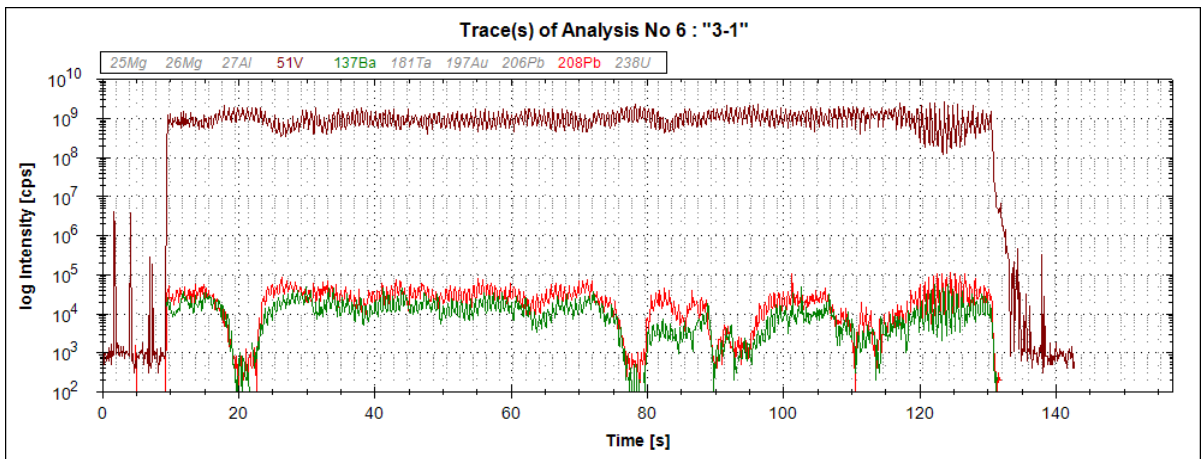

Figure S3: Sample C, line 1: The varying intensity of $^{137}$Ba and $^{208}$Pb in comparison to the constant $^{51}$V signal.

## 1.2. XPS: Comparison of monoclinic and tetragonal VO$_2$

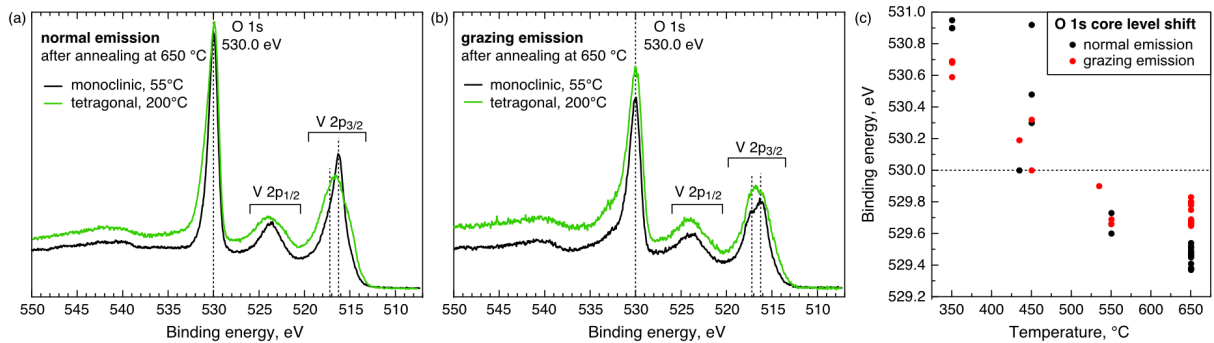

Figure S4: Normal (a) and grazing (b) emission of the monoclinic and tetragonal phases of VO$_2$(110)$_T$. The spectra were aligned to the O 1s core level positioned at 530.0 eV. (c) Original binding energies of the O 1s core level for normal and grazing emission at different annealing temperatures (compensated in panels (a, b)).

## 2. DFT: Modified ring terminations

Since the ring termination with the surface stoichiometry $V_4O_{13}$ described in the main text is only one out of many possibilities for such a structure, we also studied its modifications. The most promising candidates are described in this section. The original ring structure with the surface stoichiometry $V_4O_{13}$ can be adapted by adding another vanadium tetrahedron between the rows as is shown in Figure S5(a-b), forming $V_5O_{14}$ and $V_5O_{15}$ rings. The structure in panel (a) with the surface stoichiometry $V_5O_{14}$ connects the rings with another vanadium tetrahedron that is bound to the subsurface layer. The vanadium rings are connected in this case by a single V-O bond to the subsurface layer, as in the ring termination in the Figure 11(c). The $V_5O_{14}$ configuration is, therefore, structurally similar to the $SrTiO_3(110)$ termination that is shown as the (3×1) surface structure in ref. [S1], but with two major differences. The first difference is related to the bonding of the superstructure to the bulk termination. In the case of $VO_2$, vanadium tetrahedra are bound only with a single V-O bond to the $VO_2$ bulk, *i.e.*, this superstructure is more open compared to the titanium tetrahedra that terminate the $SrTiO_3(110)$ surface. Secondly, some vanadium tetrahedra are additionally oxidized with another oxygen atom, which results in a disconnection from the (110) surface and the formation of vanadyl bonds on top, as is seen by comparing the side views of the structures in the Figure S5.

Simulated STM images also show patterns that are close to the experimental findings. The shorter distance between the topmost oxygen atoms of the ring structure on the left side is reduced in size from 3.6 Å to 3.3 Å which causes a small overlap between the bright spots. The additional oxygen atom in the structure in panel (b) forms a weak protrusion that sits between the double rows.

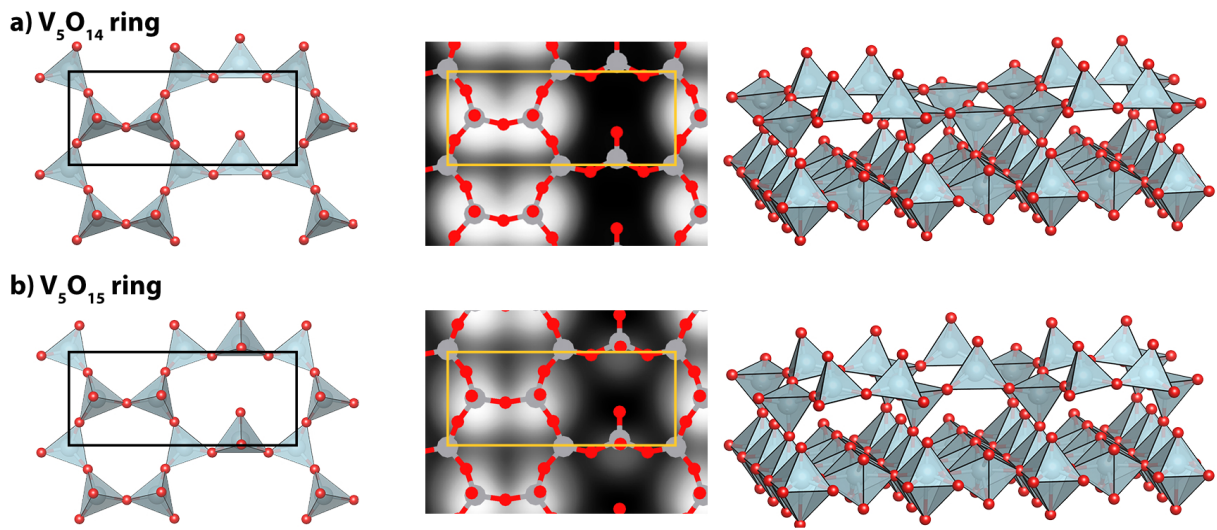

Figure S5: Top views, calculated STM images and perspective views of the relaxed modifications of two ring terminations. In panel (a) the $V_4O_{13}$ rings are connected via an additional tetrahedron that is bound to the substrate, leading to an overall $V_5O_{14}$ stoichiometry. In panel (b) one connecting tetrahedron per (2×2) cell is flipped, resulting in adsorption of an additional oxygen.